\documentclass[preprint,prd]{revtex4}
\usepackage{graphicx}
\usepackage{dcolumn}
\usepackage{bm}

\begin{document}

\title{\bf A search for the fourth SM family quarks at the Tevatron}

\author{E. Arik$^{a}$, O. \c Cak\i r$^{b}$, S. Sultansoy$^{c,d}$}
\affiliation{$a$) Bo\u{g}azi\c{c}i University, Faculty of Arts and
Sciences, Department of Physics, 80815 Bebek, Istanbul, Turkey}
\affiliation{$b$) Ankara University, Faculty of Science, Department of
Physics, 06100 Tando\u gan, Ankara, Turkey }
\affiliation{$c$) Gazi University, Faculty of Arts and  Sciences,
Department of Physics, 06500 Teknikokullar, Ankara, Turkey}
\affiliation{$d$) Azerbaijan Academy of Sciences, Institute of Physics, H.
Cavid Avenue 33, Baku, Azerbaijan}

\begin{abstract}
It is shown that the fourth standard model (SM) family quarks can be
observed at the Fermilab Tevatron if their anomalous interactions with
known quarks have sufficient strength.
\end{abstract}


\maketitle

It is known that flavor democracy \cite{1} favors the existence of
the fourth SM family fermions \cite{2,3,4}. The masses of these
fermions are expected to be nearly degenerate and lie between 300
GeV and 700 GeV. According to flavor democracy the fourth family
neutrino should be heavy. In the framework of democratic mass
matrix approach, small masses for the first three neutrinos are
compatible with large mixing angles assuming that the neutrinos
are of the Dirac type \cite{5}. Obviously, the existence of the
fourth SM family leads to a lot of cosmological and astrophysical
consequences (see for example \cite{6}).

The experimental lower bounds on the fourth SM family fermions are
as follows \cite{7}: 100.8 GeV for charged lepton, 45 (39.5) GeV
for Dirac (Majorana) neutrino and 199 (128) GeV for "down" quark
decaying via neutral (charged) current. On the other hand, the
partial-wave unitarity at high energies leads to upper limit
$m_4<1$ TeV for heavy fermions \cite{8}.

The fourth family quarks will be  copiously produced at the LHC
\cite{9,10} and the  fourth family leptons will be observed at the
future lepton colliders \cite{11,12}.

In principle, the Tevatron may also contribute to the subject.
First, the fourth family quarks can manifest themselves indirectly
due to the enhancement in the Higgs boson production \cite{13,14}.
Second, they can be produced directly via possible anomalous  $g q
q_4$ interactions. It should be noted that the arguments given in
\cite{15} for anomalous interactions of the top quark are more
valid for $u_4$ and $d_4$ quarks since they are expected to be
heavier than the top quark. In our previous papers \cite{16,17} we
have shown that the superjet events observed by the CDF
\cite{18,19,20} could be interpreted in relation to the latter
mechanism if one assumes, in addition, the existence of a new
light scalar particle, decaying dominantly to $\tau^+\tau^-$
and/or $c\bar{c}$.

In this work, we consider the anomalous production of $u_4$ and
$d_4$ quarks at the Tevatron via the subprocesses $g u(c)
\rightarrow u_4$ and $g d(s,b) \rightarrow d_4$, respectively;
followed by either SM or anomalous decays into the SM particles.

We use the following effective Lagrangian for the  anomalous
interactions of the fourth SM family quarks \cite{16}:
\begin{eqnarray}
L = \frac {\kappa_{\gamma}^{q_i}}{\Lambda} e_q g_e \bar q_4 \sigma_{\mu \nu}q_i F^{\mu \nu} +
 \frac {\kappa_Z^{q_i}}{2 \Lambda} g_Z \bar q_4 \sigma_{\mu \nu}
q_i Z^{\mu \nu}
+\,  \frac {\kappa_g^{q_i}}{\Lambda} g_s \bar q_4 \sigma_{\mu \nu}
T^a q_i G^{\mu \nu}_a + h.c.
\end{eqnarray}
where $F^{\mu \nu}, Z^{\mu \nu}$, and $G^{\mu \nu}$ are the field strength
tensors of the photon, $Z$ boson and gluons, respectively; $T^a$ are
Gell-Mann matrices;
$e_q$ is the charge of the quark;
$g_e, g_Z$, and $g_s$ are the electroweak, and
the strong coupling constants, respectively.
$g_Z = g_e/\cos\theta_W \sin\theta_W$ where $\theta_W$ is the Weinberg angle.
$\kappa_{\gamma, Z, g}^q$ define the strength of the  anomalous couplings
for the neutral currents with a photon, a $Z$ boson
and a gluon, respectively;  $\Lambda$ is the cutoff scale for the new
physics.

\begin{figure}
\includegraphics[height=7cm,width=7.5cm]{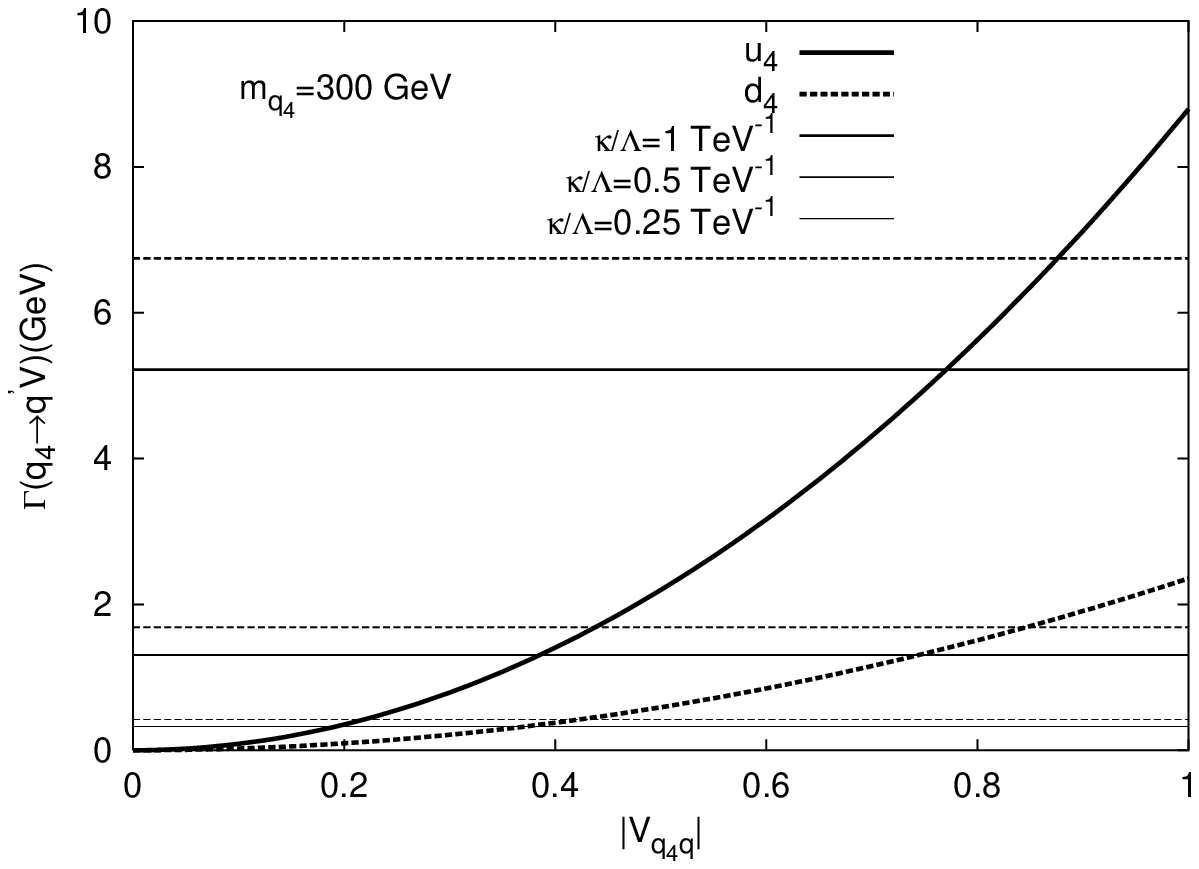}
\includegraphics[height=7cm,width=7.5cm]{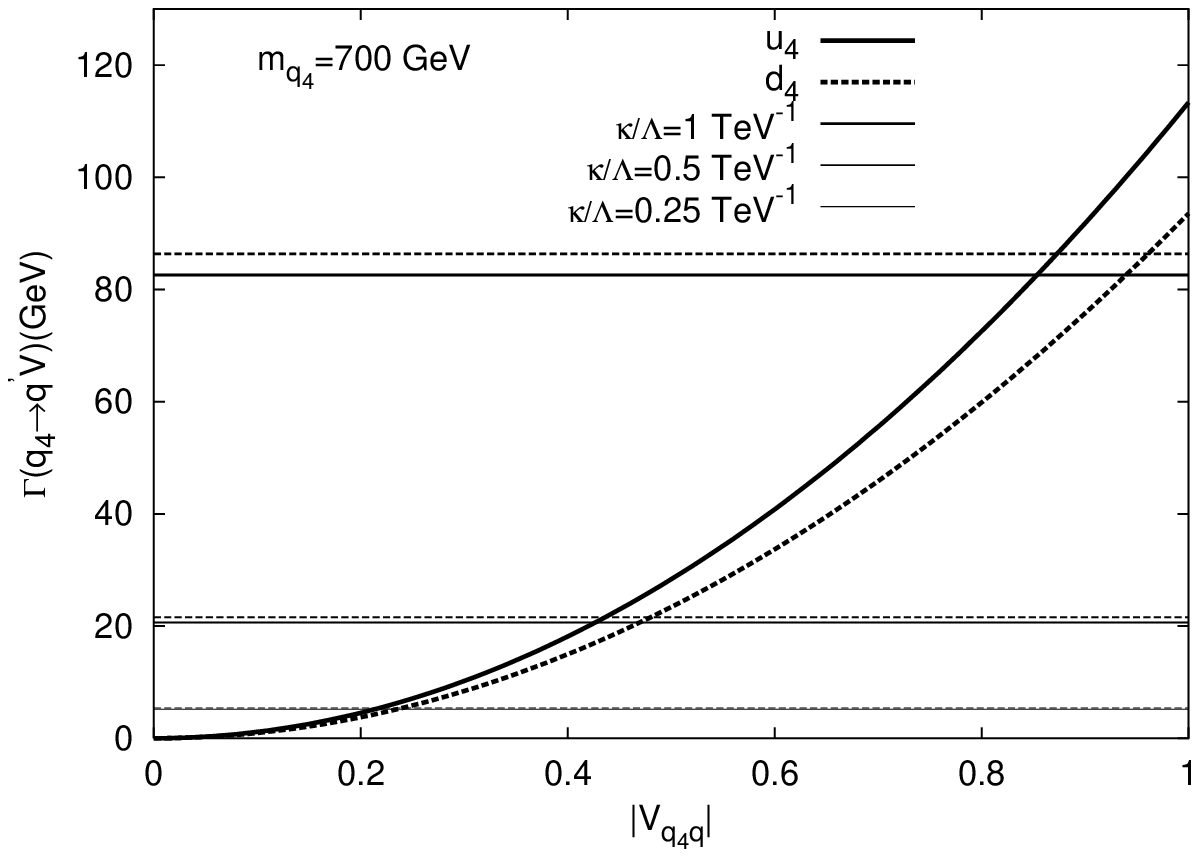}\\
\centerline{(a)$\qquad\qquad\qquad\qquad\qquad\qquad\qquad\qquad$
(b)} \caption{The decay widths for the fourth family quarks with
mass
 a) $m_{q_4}=300$ GeV and b) $m_{q_4}=700$ GeV. Horizontal lines correspond to the anomalous decay modes.}
\label{fig1}
\end{figure}

\begin{figure}
\includegraphics[height=1.5cm,width=10cm]{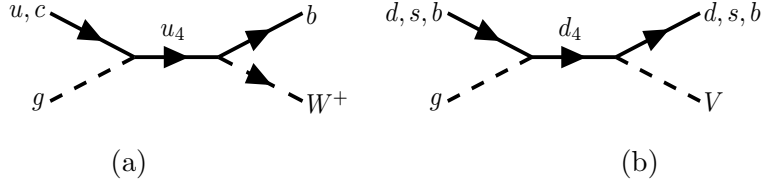}\\
\centerline{(a)$\qquad\qquad\qquad\qquad\qquad\qquad\qquad\qquad$
(b)} \caption{Anomalous production of $q_4$ quarks followed by a)
SM decay of $u_4$ quark and b) anomalous decay of $d_4$ quark
where $V=g,Z,\gamma$.} \label{fig2}
\end{figure}

In order to calculate the cross-sections and the decay widths, we
have implemented the new interaction vertices into the CompHEP
\cite{21} package. We have used the parton distribution functions
 CTEQ5L \cite{22} at $Q^2=m_{q_4}^2$. In addition to the dependence
on the fourth family quark masses, the decay widths depend on
$\kappa_V^{q_i}$ for anomalous decays and on CKM matrix elements
$V_{q_4q}$ for SM decay modes. This is demonstrated in Fig.
\ref{fig1}a(b) for $m_{q_4} = 300$ GeV ($700$ GeV). Since $u_4$
and $d_4$ quarks are almost degenerate in mass, their anomalous
$s$-channel production cross sections will be of the same order
for equal anomalous couplings. If the SM decay modes of the fourth
family quarks are dominant, an investigation of $u_4$ quark is
advantageous because of the clear $u_4\to bW^+$ signature
comparing to $d_4\to tW^-\to bW^+W^-$. If the anomolaus decay
modes are dominant $d_4$ quark has a clear signature $d_4\to bV$
($V=\gamma,Z,g$) with b-tagging. Corresponding Feynman diagrams
are shown in Fig. \ref{fig2}.

First, let us consider the process $p\bar{p}\rightarrow
u_4X\rightarrow bW^+X$. Table \ref{table1} lists the cross
sections for signal (with couplings $\kappa/\Lambda=0.5$
TeV$^{-1}$ and $0.25$ TeV$^{-1}$) and the main background
$p\bar{p}\rightarrow bW^+X$. We consider two cases: without any
cuts and with $p_T^b>50$ GeV for b-jets for two limiting $m_{u_4}$
mass values, namely, 300 GeV and 700 GeV. Numerical calculations
were performed for $|V_{u_4b}|=(\kappa/\Lambda)\cdot$TeV. This
yields the dominance of SM decay mode (see Fig. \ref{fig1}). In
order to estimate the observability limits for anomalous couplings
$\kappa/\Lambda$, we use the definition of significance
\begin{equation}
SS=\frac{\sigma_{S+B}-\sigma_B}{\sqrt{\sigma_B}}\sqrt{\epsilon\cdot
BR\cdot L_{int}}
\end{equation}
where $\epsilon=0.5$ is the detection efficiency including
b-tagging and $BR=0.2$ is the branching ratio of $W^+\rightarrow
e^+\nu_e+\mu^+\nu_\mu$. We also require the minimum number of
signal events to be $10$ and $SS\geq 5$. Assuming $L_{int}=10$
fb$^{-1}$ for integrated luminosity. We obtain following low
limits on the anomalous coupling: $\kappa/\Lambda=0.03$ TeV$^{-1}$
without cuts and $\kappa/\Lambda=0.01$ TeV$^{-1}$ with $p_T^b>50$
GeV for $m_{u_4}=300$ GeV. Corresponding numbers for $m_{u_4}=700$
GeV are $\kappa/\Lambda=0.33$ TeV$^{-1}$ and $\kappa/\Lambda=0.12$
TeV$^{-1}$, respectively.
\begin{table}
\caption{The cross sections for the process $p\bar{p}\to u_4X\to
bW^+X$ with $\kappa/\Lambda=0.5$ and $0.25$ TeV$^{-1}$, and the
corresponding background $p\bar{p}\rightarrow bW^+X$ without and
with $p_T$ cuts.} \label{table1}
\begin{tabular}{ccccc}
\hline\hline
&
\multicolumn{4}{c}{$m_{u_{4}}=300$ (700) GeV}\\
\hline
&
\multicolumn{2}{c}{no cut}&
\multicolumn{2}{c}{with cut}\\
\hline
$\kappa/\Lambda$, TeV$^{-1}$&
$0.5$&
$0.25$&
$0.5$&
$0.25$\\
\hline
$\sigma _{S+B}$, pb&
$22.1\,(0.62)$&
$5.7\,(0.21)$&
$20.4\,(0.34)$&
$5.2\,(0.082)$\\
$\sigma _{B}$, pb&
\multicolumn{2}{c}{$0.184$ }&
\multicolumn{2}{c}{$6.8\times 10^{-4}$}\\
\hline\hline
\end{tabular}
\end{table}

The next process we consider is $p\bar{p}\to d_4X\to qVX$ where
$q=d,s,b$ and $V=g,Z,\gamma$. The condition
$|V_{d_{4}t}|<(\kappa/\Lambda)\cdot$TeV ensures the anomalous
decay mode of $d_4$ to be dominant. In Table \ref{table2},
branching ratios and total decay widths of $d_4$ quark with
$m_{d_{4}}=300$ and $700$ GeV are given for the anomalous coupling
$\kappa/\Lambda=0.5$ TeV$^{-1}$ and $\kappa/\Lambda=0.25$
TeV$^{-1}$. The calculated signal cross sections for $d_4$ quark
are presented in the last row of the Table \ref{table2}.
\begin{table}
\caption{Branching ratios (BR) and decay widths $\Gamma$ for $d_4$ quark with the
anomalous coupling $\kappa/\Lambda=0.5$ and $0.25$ TeV$^{-1}$. The last
row presents the cross section for anomalous production of $d_4$ quarks.}\label{table2}
\begin{tabular}{cccccc}
\hline\hline
$m_{d_{4}}$, GeV&
&
\multicolumn{2}{c}{300}&
\multicolumn{2}{c}{700}\\
\hline
$\kappa /\Lambda ,$TeV$^{-1}$&
&
0.5&
0.25&
0.5&
0.25\\
\hline
&
$gd(s,b)$&
31&
31&
31&
31\\
BR(\%)&
$Zd(s,b)$&
1.9&
1.9&
2.1&
2.1\\
&
$\gamma d(s,b)$&
0.17&
0.17&
0.16&
0.16\\ \hline
$\Gamma $, GeV&
&
1.75&
0.44&
22.4&
5.6\\
$\sigma (p\overline{p}\rightarrow d_{4}X)$, pb&
&
21.4&
5.19&
0.18&
0.077\\
$\sigma (p\overline{p}\rightarrow b\gamma X)$, pb& &
$2.72\times10^{-3}$& & & $2.25\times 10^{-6}$\\
\hline\hline
\end{tabular}
\end{table}

Keeping in mind the assumptions for the decays of $u_4$ and $d_4$
quarks, one can differentiate between $u_4$ and $\bar{u}_4$ quarks
by identifying the charge of the lepton from $W$ decay. However,
$d_4$ and $\bar d_4$ quarks have the same final state signatures.
For this reason we will double the number of signal events in our
estimations below. The decay modes of $d_4$ and $\bar d_4$ quarks
consist of two-jet, $Z$+jet and $\gamma$+jet. Even though the
dijet mode is dominant, the extraction of the signal doesn't seem
to be promising due to the huge SM background. For the $Z$+jet
mode, again $Z\to q\bar q$ is not promising due to the large
background; $BR(Z\to l^+l^-)$ reduces the number of events a lot;
$Z\to \nu\bar\nu$ results in mono-jet final states but one cannot
reconstruct $m_{d_4}$. Hence, the optimum final state is
$\gamma$+jet. The background for this process is also large but,
it can be reduced if one uses the advantage of b-tagging. For this
reason we consider the signal process $p\bar p\to d_4X\to b\gamma
X$ with the main background $p\bar{p}\rightarrow b\gamma X$.
\begin{figure}
\includegraphics[height=9cm,width=12cm]{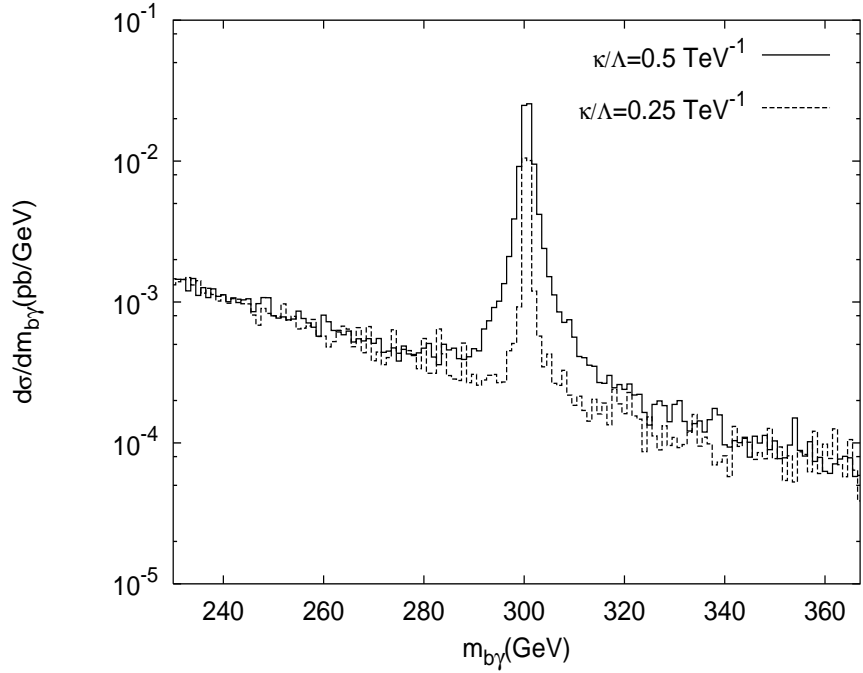}\\
\caption{Invariant mass distribution of b-tagged jet and photon for signal
($m_{d_{4}}=300$ GeV, $\kappa/\Lambda=0.5$ and $0.25$ TeV$^{-1}$) and
background.}
\label{fig3}
\end{figure}

For illustration, in Fig. \ref{fig3} we present the invariant mass
distribution of the background and signal events for $m_{d_4}=300$
GeV, and two values of $\kappa/\Lambda$ (0.5 and 0.25 TeV$^{-1}$).
Obviously, from Fig. \ref{fig3} the signal is quite observable. As
$\kappa/\Lambda$ decreases and/or $m_{d_4}$ increases the
situation gets worse. In order to observe at least 10 signal
events with $SS\geq 5$ at $L_{int}=10$ fb$^{-1}$, the anomalous
coupling should satisfy $\kappa/\Lambda\ge$ 0.08 (0.9) TeV$^{-1}$
for $m_{d_4}=300$ GeV ($m_{d_4}=700$ GeV).

In conclusion, the fourth SM family quarks could be observed at
the upgraded Tevatron depending on the anomalous coupling and the
mass values. For $L_{int}=10$ fb$^{-1}$, $u_4$ quark with mass 300
GeV and SM decay mode can be observed if $\kappa/\Lambda>0.01$
TeV$^{-1}$. For $m_{u_4}=700$ GeV, the lower limit on
$\kappa/\Lambda$ is 0.12 TeV$^{-1}$. On the other hand, $d_4$
quark with mass 300 (700) GeV and anomalous decay mode can be
observed if $\kappa/\Lambda>0.08$ (0.9) TeV$^{-1}$.

This work is partially supported by Turkish State Planning
Organization under the Grant No 2002K120250 and 2003K120190.

\end{document}